\documentstyle[12pt]{article}
\begin{document}

\author{R. Vilela Mendes\\
Grupo de F\'\i sica-Matem\'atica \\
%EndAName
Complexo Interdisciplinar, Universidade de Lisboa \\
Av. Gama Pinto, 2, 1699 Lisboa Codex Portugal\\
e-mail: vilela@alf4.cii.fc.ul.pt}
\title{Conditional exponents, entropies and a measure of dynamical
self-organization }
\date{}
\maketitle

\begin{abstract}
In dynamical systems composed of interacting parts, conditional exponents,
conditional exponent entropies and cylindrical entropies are shown to be
well defined ergodic invariants which characterize the dynamical
selforganization and statitical independence of the constituent parts. An
example of interacting Bernoulli units is used to illustrate the nature of
these invariants.
\end{abstract}

\section{Conditional exponents}

The notion of conditional Lyapunov exponents (originally called sub-Lyapunov
exponents) was introduced by Pecora and Carroll in their study of
synchronization of chaotic systems\cite{Pecora1} \cite{Pecora2}. It turns
out, as I will show below, that, like the full Lyapunov exponent, the
conditional exponents are well defined ergodic invariants. Therefore they
are reliable quantities to quantify the relation of a global dynamical
system to its constituent parts and to characterize dynamical
selforganization.

Given a dynamical system defined by a map $f:M\rightarrow M$ , with $%
M\subset R^m$ the {\it conditional exponents associated to the splitting }$%
R^k\times R^{m-k}$ are the eigenvalues of the limit 
\begin{equation}
\label{1.1}\lim _{n\rightarrow \infty }\left( D_kf^{n*}(x)D_kf^n(x)\right)
^{\frac 1{2n}} 
\end{equation}
where $D_kf^n$ is the $k\times k$ diagonal block of the full Jacobian.

{\it Lemma. Existence of the conditional exponents as well defined ergodic
invariants is guaranteed under the same conditions that establish the
existence of the Lyapunov exponents}

Proof: Let $\mu $ be a probability measure in $M\subset R^m$ and $f$ a
measure-preserving $M\rightarrow M$ mapping such that $\mu $ is ergodic.
Oseledec's multiplicative ergodic theorem\cite{Oseledec}, generalized for
non-invertible $f$ \cite{Raghu}, states that if the map $T:M\rightarrow M_m$
from $M$ to the space of $m\times m$ matrices is measurable and 
\begin{equation}
\label{1.2}\int \mu (dx)\log ^{+}\left\| T(x)\right\| <\infty 
\end{equation}
(with $\log ^{+}g=\max \left( 0,\log g\right) $) and if 
\begin{equation}
\label{1.3}T_x^n=T(f^{n-1}x)\cdots T(fx)T(x) 
\end{equation}
then 
\begin{equation}
\label{1.4}\lim _{n\rightarrow \infty }\left( T_x^{n*}T_x^n\right) ^{\frac
1{2n}}=\Lambda _x 
\end{equation}
exists $\mu $ almost everywhere.

If $T_x$ is the full Jacobian $Df(x)$ and if $Df(x)$ satisfies the
integrability condition (\ref{1.2}) then the Lyapunov exponents exist $\mu -$%
almost everywhere. But if the Jacobian satisfies (\ref{1.2}), then the $%
m\times m$ matrix formed by the diagonal $k\times k$ and $m-k\times m-k$
blocks also satisfies the same condition and conditional exponents too are
defined a. e.. Furthermore, under the same conditions as for Oseledec's
theorem, the set of regular points is Borel of full measure and 
\begin{equation}
\label{1.5}\lim _{n\rightarrow \infty }\frac 1n\log \left\|
D_kf^n(x)u\right\| =\xi _i^{(k)} 
\end{equation}
with $0\neq u\in E_x^i/E_x^{i+1}$ , $E_x^i$ being the subspace of $R^k$
spanned by eigenstates corresponding to eigenvalues $\leq \exp (\xi
_i^{(k)}) $.

\section{Conditional entropies and dynamical selforganization}

For measures $\mu $ that are absolutely continuous with respect to the
Lebesgue measure of $M$ or, more generally, for measures that are smooth
along unstable directions (SBR\ measures) Pesin's\cite{Pesin} identity holds 
\begin{equation}
\label{2.1}h(\mu )=\sum_{\lambda _i>0}\lambda _i
\end{equation}
relating Kolmogorov-Sinai entropy $h(\mu )$ to the sum of the Lyapunov
exponents. By analogy we may define the {\it conditional exponent entropies}
associated to the splitting $R^k\times R^{m-k}$ as the sum of the positive
conditional exponents counted with their multiplicity 
\begin{equation}
\label{2.2}h_k(\mu )=\sum_{\xi _i^{(k)}>0}\xi _i^{(k)}
\end{equation}
\begin{equation}
\label{2.3}h_{m-k}(\mu )=\sum_{\xi _i^{(m-k)}>0}\xi _i^{(m-k)}
\end{equation}
The Kolmogorov-Sinai entropy of a dynamical system measures the rate of
information production per unit time. That is, it gives the amount of
randomness in the system that is not explained by the defining equations (or
the minimal model\cite{Crutchfield}). Hence, the conditional exponent
entropies may be interpreted as a measure of the randomness that would be
present if the two parts $S^{(k)}$ and $S^{(m-k)}$ were uncoupled. The
difference $h_k(\mu )+h_{m-k}(\mu )-h(\mu )$ represents the effect of the
coupling.

Given a dynamical system $S$ composed of $N$ parts $\{S_k\}$ with a total of 
$m$ degrees of freedom and invariant measure $\mu $, one defines a {\it %
measure of dynamical selforganization} $I(S,\Sigma ,\mu )$ as 
\begin{equation}
\label{2.5}I(S,\Sigma ,\mu )=\sum_{k=1}^N\left\{ h_k(\mu )+h_{m-k}(\mu
)-h(\mu )\right\} 
\end{equation}
Of course, for each system $S$, this quantity will depend on the partition $%
\Sigma $ into $N$ parts that one considers. $h_{m-k}(\mu )$ always denotes
the conditional exponent entropy of the complementar of the subsystem $S_k$.
Being constructed out of ergodic invariants, $I(S,\Sigma ,\mu )$ is also a
well-defined ergodic invariant for the measure $\mu $. $I(S,\Sigma ,\mu )$
is formally similar to a mutual information. However, not being strictly a
mutual information, in the information theory sense, $I(S,\Sigma ,\mu )$ may
take negative values.

Another ergodic invariant that may be associated to the splitting of a
dynamical system into its constituent parts is the notion of {\it %
cylindrical entropies}.

Consider, as before, a $\mu -$preserving and $\mu -$ergodic mapping $%
f:M\rightarrow M$ and a splitting $R^m=R^k\times R^{m-k}$. A measure in $R^m$
induces a measure in $R^k$ by 
\begin{equation}
\label{2.6}\nu (x)=\int_{R^{m-k}}d\mu (y,x) 
\end{equation}
$x\in R^k$ and $y\in R^{m-k}$.

Given a $\nu -$measurable partition $P(R^k)$ in $R^k$%
\begin{equation}
\label{2.7}R^k=\cup _iP_i 
\end{equation}
$P_i\in $ $P(R^k)$ , it induces a partition in $R^m$ by the associated
cylinder sets 
\begin{equation}
\label{2.8}R^m=\cup _iP_i^c 
\end{equation}
$P_i^c=P_i\times R^{m-k}\in P^c(R^m)$.

Let $P^c(M)=P^c(R^m)\cap M$ be the corresponding partition of $M$. Denote by 
$P^c(x)$ the element of $P^c(M)$ that contains $x$. If all powers of $f$ are
ergodic, for any nontrivial partition 
\begin{equation}
\label{2.9}\lim _{n\rightarrow \infty }\mu (P_n^c(x))=0
\end{equation}
where 
\begin{equation}
\label{2.10}P_n^c(x)=\cap _{j=0}^nf^{-j}(P^c(f^j(x)))
\end{equation}
Then, the Shannon-MacMillan-Breiman theorem states that if 
\begin{equation}
\label{2.11}\sum_i\mu (P_i^c)\log (P_i^c)<\infty 
\end{equation}
the limit 
\begin{equation}
\label{2.12}h^c(f,P^c,x)=-\lim _{n\rightarrow \infty }\frac 1n\log \mu
(P_n^c(x))
\end{equation}
exists $\mu $ a. e. and converges in $L^1$. This limit is the entropy at $x$
associated to the cylindrical partition $P^c$. The {\it cylindrical entropy
relative to the splitting }$R^k\times R^{m-k}$ may be defined as the
integral of the supremum of this limit over all finite cylindrical
partitions 
\begin{equation}
\label{2.13}h^c(f)=-\int_Md\mu (x)\sup _{P^c}\lim _{n\rightarrow \infty
}\frac 1n\log \mu (P_n^c(x))
\end{equation}
The full Kolmogorov-Sinai entropy is a similar limit where now the supremum
would be taken over all finite partitions. Therefore for a smooth measure,
if the parts of a composite dynamical system are all uncoupled, the full
entropy is simply the sum of the cylindrical entropies. In the uncoupled
case each cylindrical entropy is determined by the corresponding conditional
exponents. However for coupled mixing systems, the cylindrical partitions
may, by themselves, already generate the full entropy of the coupled system.
Therefore the relation of the cylindrical entropies to the total entropy is
simply a {\it measure of the statistical independence} of the constituent
parts. The conditional exponent entropies defined in (\ref{2.2}-\ref{2.3})
seem to be a better quantitative characterization of the dynamical
selforganization.

\section{An example}

Consider a fully coupled system defined by 
\begin{equation}
\label{3.1}x_i(t+1)=(1-c)f(x_i(t))+\sum_{j\neq i}\frac c{N-1}f(x_j(t))
\end{equation}
with $f(x)=2x$ (mod. $1$).

The Lyapunov exponents are $\lambda _1=\log 2$ and $\lambda _i=\log \left(
2\left( 1-\frac N{N-1}c\right) \right) $ with multiplicity $N-1$.

Therefore, for an absolutely continuous measure 
\begin{equation}
\label{3.2}
\begin{array}{cclcc}
h(\mu ) & = & \log 2+(N-1)\log \left( 2-\frac{2Nc}{N-1}\right) & \textnormal{for}
& c\leq 
\frac{N-1}{2N} \\  & = & \log 2 & \textnormal{for} & c\geq \frac{N-1}{2N} 
\end{array}
\end{equation}
The conditional exponents associated to the splitting $R^1\times R^{N-1}$
are 
\begin{equation}
\label{3.3}\xi ^{(1)}=\log (2-2c) 
\end{equation}
and 
\begin{equation}
\label{3.4}
\begin{array}{ccccccccc}
\xi _1^{(N-1)} & = & \log \left( 2-\frac{2c}{N-1}\right) & ; & \xi
_i^{(N-1)} & = & \log \left( 2-\frac{2Nc}{N-1}\right) & \textnormal{with
multiplicity} & N-2 
\end{array}
\end{equation}
Therefore, for a partition $\Sigma $ of the system with $N$ parts one
obtains 
\begin{equation}
\label{3.5}I(S,\Sigma ,\mu )=N\left( \log \left( 1-\frac c{N-1}\right) +\max
\left( \log (2-2c),0\right) -\max \left( \log \left( 2-\frac{2Nc}{N-1}%
\right) ,0\right) \right) 
\end{equation}
which in the limit of large $N$ becomes 
\begin{equation}
\label{3.6}
\begin{array}{ccccc}
I(S,\Sigma ,\mu ) & = & \frac{c^2}{1-c} &  & c\leq 
\frac{N-1}{2N} \\  & = & -c &  & c\geq \frac 12 
\end{array}
\end{equation}
Fig.1 shows the variation with $c$ of $I(S,\Sigma ,\mu )$ for $N=100$.

At $c=0$, and starting from a random initial condition, the motion of the
system is completely disorganized. When $c$ starts to grow the system shows
the coexistence of disorganized behavior with patches of synchronized
clusters. At the point where $I(S,\Sigma ,\mu )$ is maximum, $c=0.495$,
starting from a random initial condition, the system settles rapidly in a
state with many different synchronized clusters. Fig.2 shows the first $5000$
time steps. It is indeed at this point that the system shows what
intuitively we would call a large organizational structure. Above $c=0.5$,
after a short transition period, the system becomes fully synchronized
(Fig.3 for $c=0.51$).

\section{Figure captions}

Fig.1 - Coupling dependence of the selforganization invariant $I(S,\Sigma
,\mu )$ in the coupled Bernoulli system

Fig.2 - The first $5000$ time steps for $c=0.495$ (maximum $I(S,\Sigma ,\mu )
$). The last column in the right is the color map

Fig.3 - The first $5000$ time steps for $c=0.51$


\begin{thebibliography}{9}
\bibitem{Pecora1}  L. M. Pecora and T. L. Carroll; Phys. Rev. Lett. 64
(1990) 821.

\bibitem{Pecora2}  L. M. Pecora and T. L. Carroll; Phys. Rev. A44 (1991)
2374.

\bibitem{Oseledec}  V. I. Oseledec; Trans. Moscow Math. Soc. 19 (1968) 197.

\bibitem{Raghu}  M. S. Raghunatan; Israel Jour. Math. 32 (1979) 356.

\bibitem{Pesin}  Y. B. Pesin; Russ. Math. Surv. 32 (1977) 55.

\bibitem{Crutchfield} J. P. Crutchfield and K. Young; in {\it Complexity, 
Entropy and the Physics of Information}, pag. 223, SFI Studies in the 
Sciences Complexity, vol. VIII, W. H. Zurek (Ed.), Addison-Wesley 1990.
\end{thebibliography}
\end{document}